\documentclass[english,aps,prc,twocolumn,amsmath,amssymb,preprintnumbers,nofootinbib,
superscriptaddress]{revtex4}

\usepackage[T1]{fontenc}
\usepackage[latin9]{inputenc}
\usepackage{babel}
\usepackage{amsmath}
\usepackage{graphicx}
\usepackage{color}
\usepackage[unicode=true]{hyperref}
\usepackage{siunitx}
\usepackage{enumerate}

\newcommand{\ga}{\gamma}

\newcommand{\beq}{\begin{equation}}
\newcommand{\eeq}{\end{equation}}
\newcommand{\ba}{\begin{array}}
\newcommand{\ea}{\end{array}}
\newcommand{\bea}{\begin{eqnarray}}
\newcommand{\eea}{\end{eqnarray}}
\newcommand{\bi}{\begin{itemize}}  %\setlength{\itemsep}{0\parsep}}
\newcommand{\ei}{\end{itemize}}
\newcommand{\ben}{\begin{enumerate}} %\setlength{\itemsep}{0\parsep}}
\newcommand{\een}{\end{enumerate}}
\newcommand{\bc}{\begin{center}}
\newcommand{\ec}{\end{center}}

\newcommand{\eqn}[1]{(\ref{#1})}

\newcommand{\MeV}{{\rm MeV}}
\newcommand{\keV}{{\rm keV}}
\newcommand{\nsat}{n_{\text{sat}}}
\newcommand{\muI}{\mu_{I}}
\newcommand{\mud}{\mu_{\delta}}
\addto\captionsenglish{}
\begin{document}
\title{Beta equilibrium in neutron star mergers}

\author{Mark G. Alford}
\affiliation{Physics Department, Washington University, St.~Louis, MO~63130, USA}
\author{Steven P. Harris}
\affiliation{Physics Department, Washington University, St.~Louis, MO~63130, USA}

\begin{abstract}
We show that the commonly used criterion for beta equilibrium in neutrino-transparent dense nuclear matter becomes invalid as temperatures rise above 1\,MeV. Such temperatures are attained in neutron star mergers. By numerically computing the relevant weak interaction rates we find that the correct criterion for beta equilibrium requires an isospin chemical potential that can be as large as 10-20\,MeV, depending on the temperature at which neutrinos become trapped.
\end{abstract}

\date{14 Sept 2018} % Hardwire date so arXiv doesn't change it

\maketitle

%%%%%%%%%%%%%%%%%%%%%%%%%%%%%%%%%%%%%%%

\section{Introduction}
\label{sec:intro}

Beta-equilibrated nuclear matter is of great physical significance
as the main constituent of neutron stars. It is therefore
important to establish the conditions for beta equilibrium at the
densities and temperatures that are astrophysically relevant.

In this paper we point out that the standard low-temperature criterion
for beta equilibrium in neutrino-transparent nuclear matter
is not valid at temperatures 
%($T$ in the range 1 to 50\,\MeV)
and densities 
% ($n$ up to a few times  $\nsat$)
that are attained in neutron star mergers. 
After their formation in a supernova, neutron stars quickly cool \cite{Yakovlev_cooling,Yakovlev_cooling_review,ns_cooling_observed}
below a temperature of 1\,MeV, but neutron star
mergers (now observed \cite{NS_NS_1,NS_NS_2})
contain nuclear matter at densities above nuclear saturation density $\nsat$ and temperatures ranging up to 30\,MeV \cite{Baiotti:2016qnr,how_loud,mass_ejection,Rezzola_original,APR_merger,shibata}.
The proper way to treat nuclear matter in mergers is to do a full neutrino transport calculation. However, this is a technically formidable task so many treatments (e.g., \cite{shibata,Endrizzi:2016kkf,Bauswein:2010dn,Hotokezaka:2011dh,Maione:2016zqz}) assume neutrino transparency up to some temperature $T_{\rm trap}$, using a cold neutrinoless beta-equilibrated equation of state for $T < T_{\rm trap}$. In this work we provide the leading finite-temperature corrections to the beta equilibration condition in neutrino-transparent matter.
We find that beta equilibrium requires an isospin chemical potential that increases with the temperature $T$, reaching a few MeV at $T\sim 1\,\MeV$ and
rising to almost 25\,MeV at $T\sim 10\,\MeV$.
Since the neutron stars start the merger process in a state of high density but low temperature, and then heat up as the merger proceeds,
we expect that there will be regions in which neutrinos are not trapped
and the temperature and density are in the range where our predicted corrections
are significant.  The size of such ranges is not yet clear because current estimates of the neutrino mean free path
depend on the energy of the neutrinos and on the
assumptions about nuclear matter \cite{neutrino_trapping,particle_hole_excitations,roberts_reddy,nuclear_mean_field}.  In Appendix \ref{sec:MFP} we summarize a recent calculation of the mean free path of neutrinos, indicating that neutrinos are not trapped until temperatures rise above approximately $5\,\MeV$.

%We perform numerical calculations of the rates of the 
%relevant flavor-changing processes at these densities and temperatures
%and show that in nuclear matter,
%beta equilibrium requires an isospin chemical potential that rises with the 
%temperature $T$, reaching almost 30\,MeV at $T\sim 20\,\MeV$.

We work in natural units, where $\hbar=c=k_B=1$.
%%%%%%%%%%%%%%%%%%%%%%%%%%%%%%%
\subsection{The Fermi surface approximation}

Neutron stars contain nuclear matter at temperatures that are
much smaller than the Fermi energy of the nucleons and electrons.
Neutron stars are finite-size systems so we also have to
take into account finite size effects such as neutrino transparency,
which occurs at temperatures low enough so that
the neutrino mean free path is larger than the size
of the star \cite{shapiro,Yakovlev_review,roberts_reddy,neutrino_trapping}.
At high enough density, flavor equilibration occurs
via direct Urca processes (neutron decay and electron capture)
\beq
\ba{rcl}
n &\rightarrow& p+ e^- + \bar\nu_e \ ,\\
p + e^- &\rightarrow& n+ \nu_e \ .
\ea
\label{eq:Urca}
\eeq
Note that because neutrinos escape from the star, 
neutrinos can only occur in the final state.

%It is usually assumed that at such temperatures obeying
%conditions (a) and (b)
In the $T\to 0$ limit,
one can work in the ``Fermi surface (FS) approximation'' where
the Urca processes are dominated by 
nucleons and electrons close to their Fermi surfaces.
In the Fermi surface approximation, the
criterion for beta equilibrium in nuclear matter is
\beq
\mu_n = \mu_p + \mu_e,
\label{eq:beta}
\eeq
which can be obtained by ignoring the neutrinos in the Urca process and
thinking of the Urca reactions as $n\leftrightarrow p + e^-$, in which case
the principle of detailed balance tells us that \eqn{eq:beta} is the 
condition for equilibrium.  Here $\mu_n$, $\mu_p$, and $\mu_e$ are the relativistic chemical potentials (i.e., including the rest mass) of the neutron, proton, and electron, respectively.

In the Fermi surface approximation there is a
threshold density above which
the direct Urca processes shown in \eqn{eq:Urca} are dominant.
Below that threshold they are kinematically forbidden, and
equilibration occurs via modified Urca processes which
involve an additional spectator nucleon in the initial and final state
\cite{Yakovlev_review,Lattimer_direct_urca,mod_urca_NS}.  

In this paper, we consider only the six Urca processes (two direct and four modified) where the charged lepton involved is an electron. Weak interactions involving positrons are negligible, since the electron chemical potential is always above $100\,\MeV$ for the densities that we will consider, and so the positron occupation is suppressed by a factor of more than $\exp(-100\,\MeV/T)$.  Additionally, for simplicity we neglect Urca processes involving muons, because even though those processes are not negligible, they do not qualitatively change the conclusions that we present here.
%%%%%%%%%%%%%%%%%%%%%%%%%%%%%%%
\subsection{Breakdown of the Fermi surface approximation}

The beta equilibrium criterion \eqn{eq:beta} is not 
guaranteed to be valid under
all circumstances because neutrinos are not in statistical equilibrium
in systems that are neutrino-transparent.
Because neutrinos escape, they only occur
in final states \footnote{
The fact that neutrinos
occur only in final states means that they are out of
statistical equilibrium. This is different from being in statistical equilibrium
 with zero chemical potential, in which case there would be a thermal
population of neutrinos which could occur in initial
states as well.}
so the two processes \eqn{eq:Urca} whose rates have to balance
are not exact inverses of each other, which means that the principle
of detailed balance is not guaranteed to hold. (For a discussion
in the context of hot plasmas see Ref.~\cite{yuan_beta_eq,Liu_npe}).
Detailed balance is a good approximation when neutrinos play a
negligible role, namely in the $T\to 0$ limit where
the Fermi surface approximation is valid. However, for
astrophysical applications
we need to know the criterion for beta equilibration at a range of
temperatures up to $30\,\MeV$. 
We will now make a rough estimate of the temperature at which the corrections to Eq.~\eqn{eq:beta} become significant. In Sec.~\ref{sec:exact} we will
perform a full calculation.

For densities below the direct Urca threshold we can estimate
the range of validity of the Fermi surface approximation 
by noting that it will become invalid when
the exponential suppression of direct Urca processes involving
particles away from their Fermi surface is not so severe as to
make those processes negligible relative to modified Urca. 
In direct Urca processes
the proton is expected to play a crucial role, since it is the
most non-relativistic fermion which means that the energy $E_p$
of a proton rises very slowly as the momentum $p_p$ of the proton
deviates from its Fermi surface: $E_p-E_{F_p}\sim (p_p-p_{F_p})p_{F_p}/m_p$, where $p_{F_p}$ is the Fermi momentum of the proton, $E_{F_p}$ is the proton Fermi energy, and $m_p$ is the proton mass.
For particles on their Fermi surfaces, the momentum
mismatch for direct Urca at densities around $3\nsat$ in nuclear matter
described by the Akmal-Pandharipande-Ravenhall (APR) equation of state \cite{APR} is
$p_{\text{miss}}=p_{Fn}-p_{Fp}-p_{Fe}\approx 50\,\MeV$
(see, e.g., Fig.~2 in Ref.~\cite{Alford_transport}),
and the proton Fermi momentum is about 220\,\MeV.
The energy cost of finding a proton that is  $p_{\text{miss}}$
from its Fermi surface
is $p_{\text{miss}} p_{Fp}/m_p \approx 12\,\MeV$, so we might expect
that direct Urca electron capture, 
where the probability of finding a proton from above its Fermi surface
includes a Boltzmann factor, becomes unsuppressed at temperatures of order 10\,\MeV,
and that it starts to compete with modified Urca at even lower temperatures.  

In fact, as can be shown from the Urca rate expressions that we review in Sec.~\ref{sec:FS-approx}, at $3\nsat$ the modified Urca rate is approximately a factor of $\left(m_n T/(3 m_{\pi}^2) \right)^2$ smaller than the above-threshold direct Urca rate, where $m_n$ and $m_{\pi}$ are the neutron and pion masses, respectively.  Thus, the below-threshold direct Urca electron capture rate would begin to compete with modified Urca when
\begin{equation}
e^{-(E_p-E_{F_p})/T} \approx \left( \dfrac{m_n T}{3m_{\pi}^2} \right)^{\!2},
\end{equation}
which, for a proton with $E_p-E_{F_p} = 12\,\MeV$, is when the temperature is between 1 and 2 \,MeV.

As we will show below in an explicit calculation, this is a 
fair estimate.
The Fermi surface approximation starts to become invalid
at temperatures $T\gtrsim 1\,\MeV$ which are still 
much less than the Fermi energies, and may still be low enough  
for neutrinos to escape from the star.
This leads us to expect corrections to the low-temperature
criterion \eqn{eq:beta} for beta equilibrium 
at temperatures and densities that are relevant for
neutron star mergers, in which nuclear matter is heated to temperatures 
up to 30\,\MeV.

In Sec.~\ref{sec:FS-approx} we reproduce the standard calculation of the 
rates of neutron decay and electron capture, which uses the
Fermi surface approximation. To describe nuclear matter we will use the
APR equation of state \cite{APR}.

In Sec.~\ref{sec:beyond-FS} we describe the kinematics of the below-threshold direct Urca process, and how particles away from their Fermi surface can participate in the processes, leading to exponential suppression of the direct Urca rates for temperatures below about 10~\MeV.

In Sec.~\ref{sec:exact} we describe the results of a numerical calculation
of the rates which includes contributions from the whole phase space.

%%%%%%%%%%%%%%%%%%%%%%%%%%%%%%%%%%%%%%%%%%%%%%
\section{Urca processes in the Fermi surface approximation}
\label{sec:FS-approx}

We now obtain the standard expressions for
the rate of the direct and modified Urca processes in matter with the APR equation of state. 
We will assume ultra-relativistic electrons and neutrinos, but nucleons that are non-relativistic, with dispersion relation
\begin{equation}
\label{eq:dispersion}
E_i = m_{\text{eff},i} + \dfrac{p_i^2}{2m_i}
\end{equation}
where, following Roberts \textit{et al.} \cite{nuclear_mean_field}, at each density $m_{\text{eff},i}$ is chosen such that the Fermi energy $E_{F,i} \equiv E_i(p_{F_i})$ matches the chemical potential $\mu_i$ from the APR equation of state, which is a simple way of taking into account the nuclear mean field.  For the kinetic mass $m_i$ we use the rest mass in vacuum.
%%%%%%%%%%%%%%%%%%%%%%%%%%%%%%%%%

\subsection{Direct Urca}
In the $T\to 0$ limit,
the Fermi surface approximation is valid and
conservation of energy and momentum
ensures that the direct Urca process \eqn{eq:Urca}
can only occur above the direct Urca threshold density where the triangle condition
\beq
p_{Fn} < p_{Fp} + p_{Fe} \ ,
\label{eq:dU-threshold}
\eeq
holds.  For densities below the threshold density, the electron and proton Fermi momenta are not large enough to add up to the neutron Fermi momentum (the three momentum vectors can not be made to form a triangle) and so the direct Urca reaction is prohibited.  As density increases, the proton and electron Fermi momenta grow more quickly than the neutron Fermi momentum and when the threshold density is reached, they, when co-aligned, add up to exactly the neutron Fermi momentum.  Above threshold, the proton and electron Fermi momenta can add up to the neutron Fermi momentum even when they are not co-aligned \cite{Yakovlev_review}.

The rates of the two direct Urca processes are given by \cite{Yakovlev_review,non_beq}
\begin{align}
\Gamma_{dU,nd} &= \int \dfrac{\mathop{d^3p_n}}{\left(2\pi\right)^3}\dfrac{\mathop{d^3p_p}}{\left(2\pi\right)^3}\dfrac{\mathop{d^3p_e}}{\left(2\pi\right)^3}\dfrac{\mathop{d^3p_{\nu}}}{\left(2\pi\right)^3}\langle\vert\mathcal{M}\vert^2\rangle \label{eq:ndecay}\\
&\times\left(2\pi\right)^4\delta^4(p_n-p_p-p_e-p_{\nu}) f_n\left(1-f_p\right)\left(1-f_e\right)\nonumber \\[2ex]
\Gamma_{dU,ec} &= \int \dfrac{\mathop{d^3p_n}}{\left(2\pi\right)^3}\dfrac{\mathop{d^3p_p}}{\left(2\pi\right)^3}\dfrac{\mathop{d^3p_e}}{\left(2\pi\right)^3}\dfrac{\mathop{d^3p_{\nu}}}{\left(2\pi\right)^3}\langle\vert\mathcal{M}\vert^2\rangle \label{eq:ecapture}\\
&\times\left(2\pi\right)^4\delta^4(p_n-p_p-p_e+p_{\nu}) \left(1-f_n\right)f_pf_e\nonumber,
\end{align}
where $f_i$ are the Fermi-Dirac distributions for $n,p,\text{ or }e$, and the matrix element is 
\begin{equation}
\langle\vert\mathcal{M}\vert^2\rangle = 2G^2\left(1+3g_A^2+\left(1-g_A^2\right)\dfrac{\mathbf{p}_e\cdot\mathbf{p}_{\nu}}{E_eE_{\nu}}\right),
\end{equation}
where $G^2 = G_F^2\cos^2{\theta_c} = 1.1\times10^{-22}\,\MeV^{-4}$, where $G_F$ is the Fermi coupling constant and $\theta_C$ is the Cabibbo angle, and the axial vector coupling constant $g_A = 1.26$.  
These rate integrals are evaluated in the Fermi surface approximation, so we set $\vert \mathbf{p}_i\vert = p_{F_i}$ in all smooth functions of momentum in the integral, and the neutrino three-momentum is neglected. In this approximation we can perform phase space decomposition, splitting the rate integral into an angular integral and an energy integral which can be straightforwardly evaluated revealing that the direct Urca neutron decay and electron capture rates are identical when Eq.~\eqn{eq:beta} holds, and are given by \cite{dU_bulk,angular_integrals,Yakovlev_review}
\begin{align}
\Gamma_{dU,nd} &= \Gamma_{dU,ec} =A_{dU} G^2\left(1+3g_A^2\right) m_nm_pp_{Fe}\vartheta_{dU}T^5\\
\vartheta_{dU} &\equiv \left\{
 \ba{ll}
  0 & \text{if}\ p_{Fn}>p_{Fp}+p_{Fe}\\
 1 & \text{if}\  p_{Fn}<p_{Fp}+p_{Fe},
 \ea \right. \nonumber \\[2ex]
A_{dU} &\equiv 3\left(\pi^2\zeta(3)+15\zeta(5)\right)/(16\pi^5)\approx 0.0170\nonumber \ .
\end{align}

%%%%%%%%%%%%%%%%%%%%%%%%%%%%%%%%%%%%%%

\subsection{Modified Urca}
Below the direct Urca threshold density, modified Urca processes provide the
leading contribution in the $T\to 0$ limit because, although they are
suppressed by a higher power of $T$, they are kinematically allowed
for particles on their Fermi surfaces.
Using the Fermi surface approximation
and neglecting the neutrino three-momentum 
we can perform phase space decomposition and calculate the rates.

For the modified Urca processes, which involve strong interactions between the nucleons, we use the matrix elements given by Yakovlev \textit{et al.} \cite{Yakovlev_review} and Friman and Maxwell \cite{friman_maxwell}, which involve a long-range one-pion exchange interaction.  The matrix element for neutron decay and electron capture with a neutron spectator (n-spectator modified Urca) is given by
\begin{equation}
\langle\vert\mathcal{M}_n\vert^2\rangle = 84G^2\dfrac{f_{\pi NN}^4}{m_{\pi}^4}\dfrac{g_A^2}{E_e^2}\dfrac{p_{Fn}^4}{\left(p_{Fn}^2+m_{\pi}^2\right)^2} \ ,
\end{equation}
and the matrix element for neutron decay and electron capture with a proton spectator (p-spectator modified Urca) is given by
\begin{equation}
\langle\vert\mathcal{M}_p\vert^2\rangle = 96G^2\dfrac{f_{\pi NN}^4}{m_{\pi}^4}\dfrac{g_A^2}{E_e^2}\dfrac{(p_{Fn}-p_{Fp})^4}{\bigl((p_{Fn}-p_{Fp})^2+m_{\pi}^2\bigr)^2} \ ,
\end{equation}
with the p-wave $\pi N$ coupling constant $f_{\pi NN} \approx 1$.

When the traditional beta equilibrium condition (\ref{eq:beta}) is used, the n-spectator modified Urca neutron decay and electron capture rates are equal and given by  \cite{Yakovlev_review,angular_integrals,shapiro,mU_bulk}
\begin{align}
\label{eq:mUrca_n}
& \Gamma_{\text{mU,n}}
 = A_{mU}G^2f_{\pi NN}^4g_A^2\dfrac{m_n^3m_p}{m_{\pi}^4} \dfrac{p_{Fn}^4p_{Fp}}{\left(p_{Fn}^2+m_{\pi}^2\right)^2} \, \vartheta_n  \, T^7 \ , \\[2ex]
\vartheta_n &\equiv \left\{
 \ba{ll}
  1 & \text{if}\ p_{Fn}>p_{Fp}+p_{Fe}\\
 1-\dfrac{3}{8}\dfrac{(p_{Fp}+p_{Fe}-p_{Fn})^2}{p_{Fp}p_{Fe}} & \text{if}\  p_{Fn}<p_{Fp}+p_{Fe}.
 \ea \right. \nonumber
\end{align}
See Sec.~6 of \cite{angular_integrals} for a comprehensive discussion of the integrals involved in the Fermi surface approximation of the modified Urca rates, including clarification of errors and omissions in the literature.

The p-spectator modified Urca neutron decay and electron capture rates are equal to each other when (\ref{eq:beta}) holds, and are given by \cite{Yakovlev_review,angular_integrals,shapiro,mU_bulk}
\begin{align}
\label{eq:mUrca_p}
\Gamma_{\text{mU,p}} &= \dfrac{A_{mU}}{7}G^2f_{\pi NN}^4g_A^2 \dfrac{m_nm_p^3}{m_{\pi}^4} \nonumber \\
&\times \dfrac{p_{Fn} (p_{Fn}\!-\!p_{Fp})^4}{\bigl((p_{Fn}\!-\!p_{Fp})^2+m_{\pi}^2\bigr)^2}\, \vartheta_p  \, T^7
\end{align}
\begin{align}
\vartheta_p &\equiv \left\{
\ba{ll}
\qquad 0 &\text{if} \ p_{Fn}>3p_{Fp}+p_{Fe}\\[3ex]
 \dfrac{(3p_{Fp}+p_{Fe}-p_{Fn})^2}{p_{Fn}p_{Fe}} & \text{if}\ \ba{l} p_{Fn}>3p_{Fp}-p_{Fe}\\p_{Fn}<3p_{Fp}+p_{Fe}\ea \\[3ex]
 4\dfrac{3p_{Fp}-p_{Fn}}{p_{Fn}} 
   & \text{if}\ \ba{l} 3 p_{Fp}-p_{Fe}>p_{Fn} \\ p_{Fn} >p_{Fp}+p_{Fe}\ea \\[3ex]
\Bigl( 2+3\dfrac{2p_{Fp}-p_{Fn}}{p_{Fe}} \\
 \quad -\,3\dfrac{(p_{Fp}-p_{Fe})^2}{p_{Fn}p_{Fe}}\Bigr) & \text{if}\ p_{Fn}<p_{Fp}+p_{Fe} \ .
\ea\right. \nonumber
\end{align}
where $A_{mU} \approx 7\times 2300 / (64\pi^9)\approx .0084. $  We see that the p-spectator modified Urca process does have a threshold density, in this case the density where $p_{F_n}=3p_{F_p}+p_{F_e}$, which occurs at a proton fraction $x_p = 1/65$.  Thus, the p-spectator modified Urca process is only prohibited at extremely low densities \cite{Yakovlev_review}, well below nuclear saturation density, which is the minimum density that we consider here.  
In Appendix \ref{sec:out_of_equil_mU}, we give the Fermi-surface approximation for the modified Urca rates when Eq.~(\ref{eq:beta}) is violated by an amount $\xi = (\mu_n-\mu_p-\mu_e)/T$.

In the $T\to 0$ limit, where the Fermi surface approximation is valid,
the standard low-temperature beta equilibrium condition holds: when
\eqn{eq:beta} is obeyed, the neutron decay and electron capture rates balance for both direct and modified Urca processes.

In the upper panels of Figs.~\ref{fig:dU_500keV} and \ref{fig:dU_5MeV}, we have plotted, among other curves that we explain in Sec.~\ref{sec:exact}, the Fermi-surface approximation of the two direct Urca (in dotted, green) and four modified Urca (labeled ``mU'', in blue) rates in APR matter for T=500\,\keV\ and 5\,\MeV\ respectively.  For the APR equation of state the direct Urca threshold density is around $5\nsat$.  Above threshold, the direct Urca neutron decay and electron capture rates are identical and dominate over the modified Urca processes which have no threshold in the density range we consider.  Below threshold, neither direct Urca process is allowed and so the four modified Urca processes dominate.  The two n-spectator modified Urca processes are slightly more important than the two p-spectator modified Urca processes.  As long as the Fermi surface approximation is used, and Eq.~\eqn{eq:beta} is imposed, the proton-producing Urca processes balance the neutron-producing Urca processes exactly at all densities and temperatures.

%%%%%%%%%%%%%%%%%%%%%%%%%%%%%%%%%%%%%%%%%%%

\section{Urca processes beyond the Fermi surface approximation}
\label{sec:beyond-FS}
%%%%%%%%%%%%%%%%%%%%%%%%%%%%%%%%%%%
\subsection{Particles away from their Fermi surface}

\begin{figure}
\includegraphics[width=\hsize]{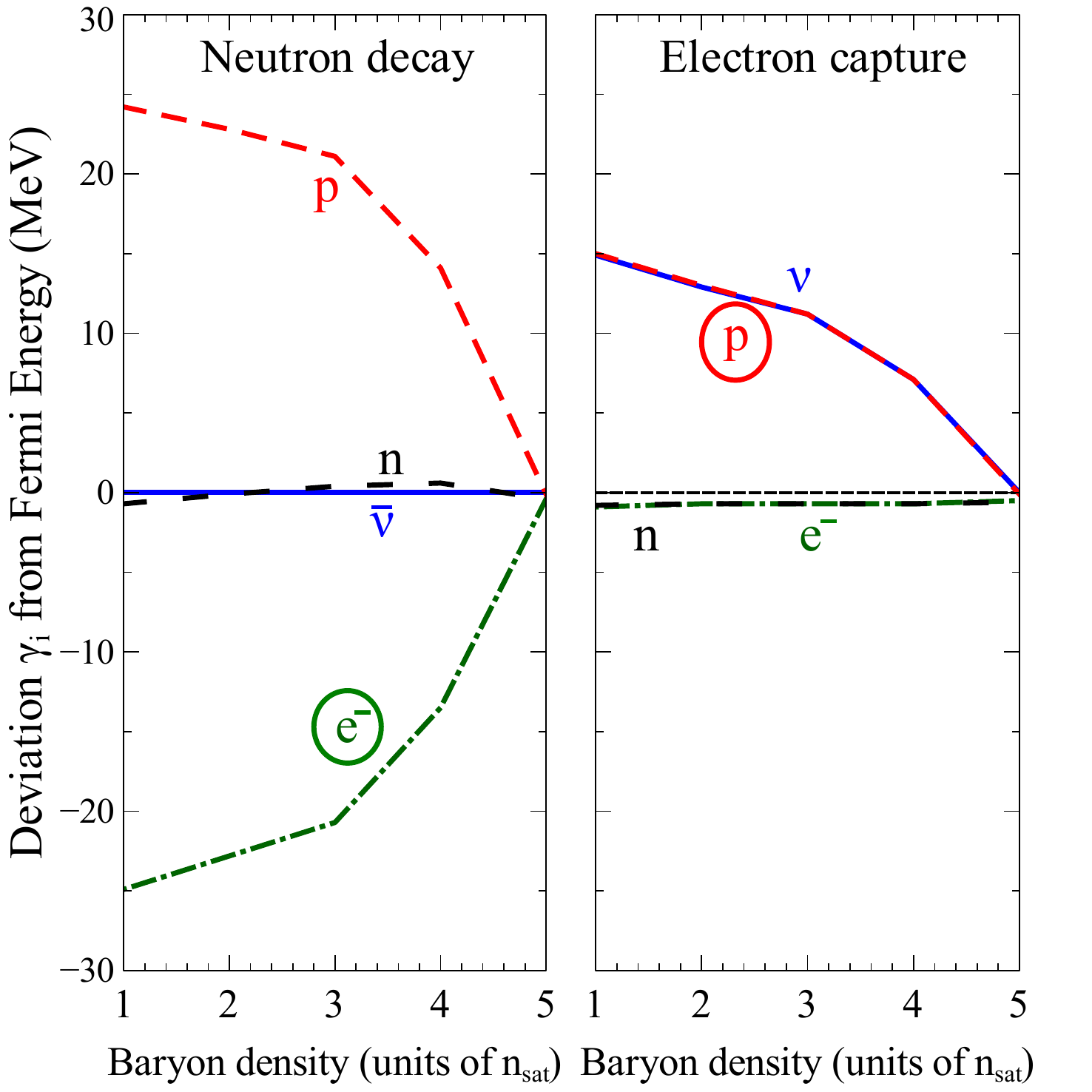}
\caption{Energy relative to their Fermi energy (defined as ``$\ga$'')
for particles participating in direct Urca reactions in APR nuclear matter
obeying the standard low-temperature condition \eqn{eq:beta} for beta equilibrium.
At each density we choose the momenta and energies of participating particles (consistent with energy and momentum conservation) 
that maximizes the product of their Fermi-Dirac factors.
Above threshold, all particles can have $\ga=0$. Below threshold, the 
plot shows the least Boltzmann-suppressed processes.  The circles indicate the particles that cause the Boltzmann suppression.
}
\label{fig:FS_deviation}
\end{figure}

To discuss the rates it is useful to introduce the
concept of the single particle free energy, defined as
$\ga_i(p) \equiv E_i(p) - \mu_i = E_i(p) - E_{F\!,i}$ (see Eq.~\ref{eq:dispersion} and subsequent discussion).
The single particle free energy tells us how far in energy a given state
is from its Fermi surface.

At densities below the threshold density, the direct Urca process becomes
Boltzmann suppressed because after imposing
energy and momentum conservation
the phase space integral is dominated by processes whose
initial state includes particles above their Fermi surface or whose 
final state requires holes below their Fermi surface. In both cases
the Fermi-Dirac factors in the rate expression provide
a suppression factor of $\exp(-|\ga_i|/T)$.

To see how strong the resultant Boltzmann suppression will be, we show
in Fig.~\ref{fig:FS_deviation} the typical single particle free energy
$\ga_i$ for the particles participating in neutron decay (left panel) and
electron capture (right panel) at various densities of nuclear matter
described by the APR equation of state, and obeying the 
low-temperature criterion
for beta equilibration \eqn{eq:beta}.
To obtain the typical momenta and energies
at a given density we impose energy and momentum conservation to reduce the momentum space integral to the lowest possible dimension and find the point at which the product of Fermi-Dirac factors attains its maximum value.  We emphasize that these typical momenta and energies are independent of temperature.  Temperature merely influences the strength of the Boltzmann suppression due to particles with finite single particle free energies $\gamma_i$ participating in Urca reactions. 

Above the direct Urca threshold density (about $5\nsat$ for the APR equation of state) we
find, as expected, that particles on their Fermi surface (i.e.~with $\ga=0$) can participate in direct Urca processes while conserving energy and momentum. Below the direct Urca threshold density, however, this is no longer true.

%%%%%%%%%%%%%%%%%%%%%%%%%%%%%%%%%%%%
\subsection{Below-threshold direct Urca neutron decay}

For direct Urca neutron decay, 
the kinematic obstacle is that although
a neutron on its Fermi surface has 
the same free energy  as a proton on its Fermi surface and an electron on its Fermi surface (they all have $\ga=0$), the neutron's momentum is larger
than the co-linear sum of the proton and electron momenta.
We see in Fig.~\ref{fig:FS_deviation} (left panel) that
the best available option below threshold
is for a neutron on its Fermi surface to decay into a proton that is
above its Fermi surface by an amount $\ga_p$ and an electron that is below its Fermi surface by the same amount, $\ga_e=-\ga_p$.
The energies of the proton and electron still add up to the energy of the
neutron, but a co-linear proton and electron now have more momentum 
then when they were both on their Fermi surfaces because
the proton's momentum rises rapidly as $\ga_p$ becomes more positive (because the proton is non-relativistic with a low Fermi velocity) whereas the electron's momentum drops more slowly as $\ga_e$ becomes more negative, because the electron is relativistic.
This ``best available option'' has a Boltzmann suppression factor of
$\exp(-|\ga_e|/T)$ because the final state electron is trying to occupy a state
in the already mostly occupied electron Fermi sea.
From Fig.~\ref{fig:FS_deviation} we see that for the APR
equation of state the value of $|\ga_e|$ for this process is
around $20$ to $25\,\MeV$ at lower
densities and then drops quickly to zero as we approach the direct Urca threshold.

%%%%%%%%%%%%%%%%%%%%%%%%%%%%%%%%%%%%%%%%
\subsection{Below-threshold direct Urca electron capture}

For direct Urca electron capture,
the kinematic obstacle is that a proton on its Fermi surface combined with
an electron on its Fermi surface does not have enough momentum to produce a
neutron on its Fermi surface.
We see in Fig.~\ref{fig:FS_deviation} (right panel) that
the best available option below threshold
is for a proton above its Fermi surface to combine with an electron at
its Fermi surface. Because the proton is nonrelativistic
this combination has enough momentum to create a neutron on its Fermi surface,
and the excess energy $\ga_p$ (and the remaining momentum) goes in to 
the final state neutrino.
This process has a Boltzmann suppression factor of
$\exp(-|\ga_p|/T)$ because we are unlikely to find initial state protons
far above the proton Fermi surface.
From Fig.~\ref{fig:FS_deviation} we see that for the APR
equation of state the value of $|\ga_p|$ for this process is
around 10 to $15\,\MeV$ at lower
densities and then drops to zero as we approach the direct Urca threshold.
The suppression is less than for neutron decay because
the neutrino momentum can be directed opposite to the neutron momentum,
so it helps to reduce the amount by which the proton needs to be above
its Fermi surface.

\begin{figure*}
\centering
\begin{minipage}{\columnwidth}
\includegraphics[width=\hsize]{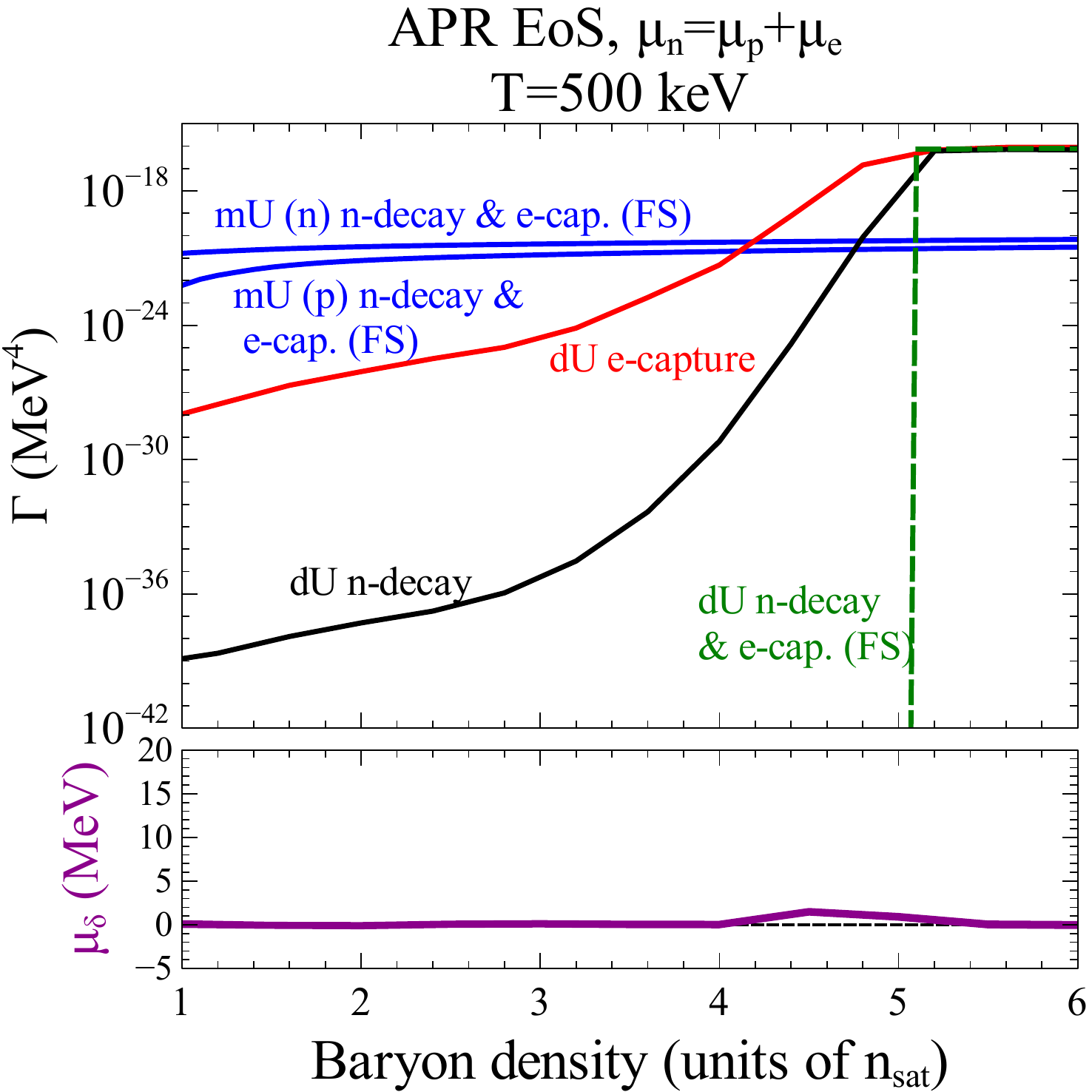}
\caption{Exact direct Urca and approximate modified Urca rates at $T = 500\,\text{keV}$, obeying the low-temperature beta equilibrium condition (\ref{eq:beta}).  Above threshold, the two direct Urca rates balance each other, and they also match the approximate direct Urca rate.  Below threshold, the direct Urca rates exponentially fall off, and modified Urca dominates.  However, the two direct Urca rates have different exponential falloffs below threshold.  In the lower plot, the deviation $\mud$ from low-temperature beta equilibrium needed to achieve true beta equilibrium is plotted as a function of density.  
}
\label{fig:dU_500keV}
\end{minipage}\hspace{.02\textwidth}
\begin{minipage}{\columnwidth}
\includegraphics[width=\hsize]{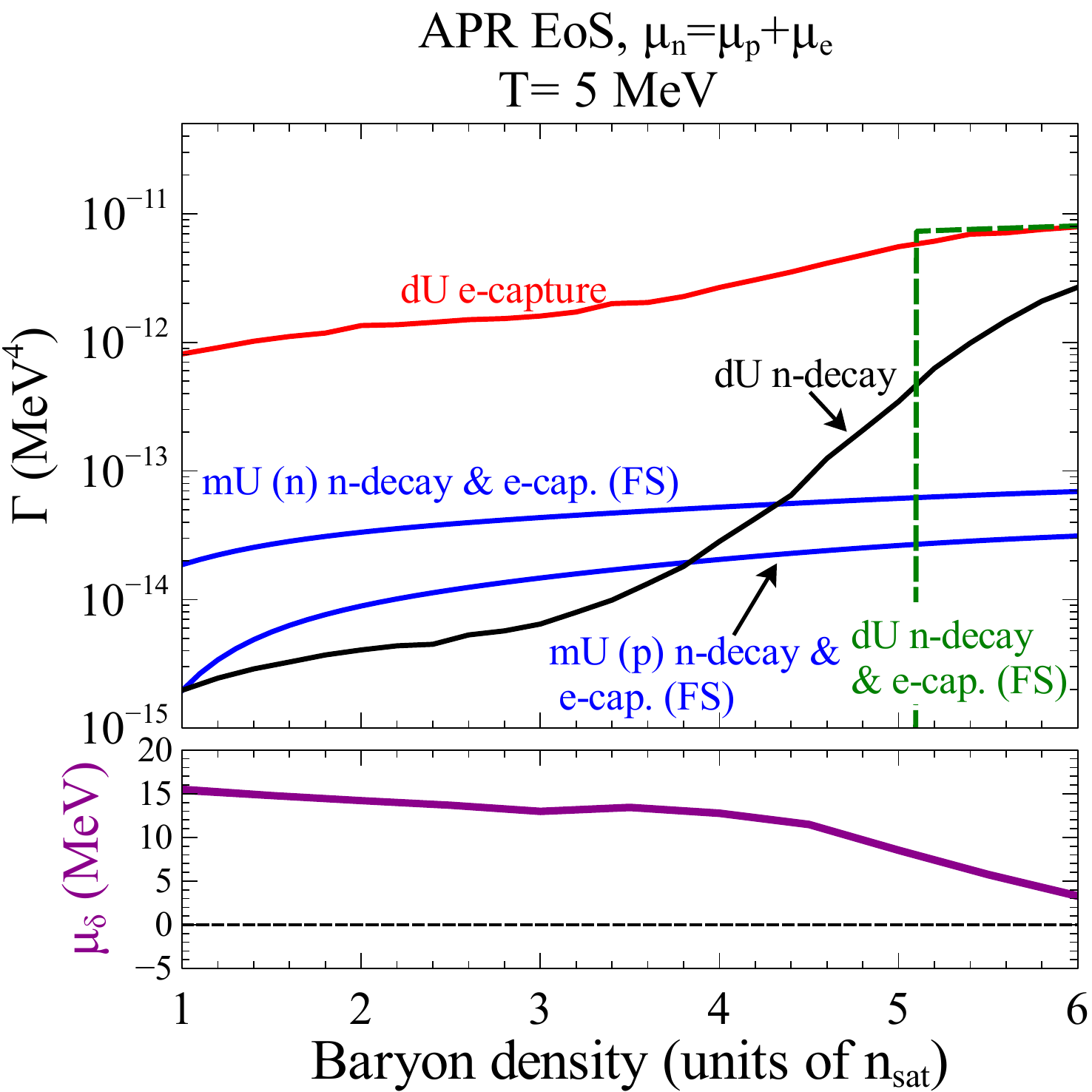}
\caption{Exact direct Urca and approximate modified Urca rates at $T = 5\,\MeV$, obeying the low-temperature beta equilibrium condition (\ref{eq:beta}).  Above threshold, electron capture direct Urca dominates, and agrees with the approximate direct Urca rate calculation.  At densities greater than $6\nsat$, we expect neutron decay direct Urca to match electron capture direct Urca.  Below threshold, electron capture direct Urca dominates over all modified Urca processes, which contradicts the conventional wisdom.  In the lower plot, the deviation $\mud$ from low-temperature beta equilibrium needed to achieve true beta equilibrium is plotted as a function of density.   
}
\label{fig:dU_5MeV}
\end{minipage}
\end{figure*} 
%%%%%%%%%%%%%%%%%%%%%%%%%%%%%%%%%%%%%%%%%%%%
\subsection{Relevance of below-threshold direct Urca}

We learn from the calculation presented in Fig.~\ref{fig:FS_deviation} that,
below the threshold density, direct Urca processes are Boltzmann suppressed by
a factor $\exp(-\ga/T)$ where $\ga$ is in the 10 to 20 \,MeV range at lower
densities, dropping to zero as the threshold is approached. For typical
neutron star temperatures $T\lesssim 0.1\,\MeV$ the Boltzmann suppression is
overwhelming, and direct Urca processes can be safely neglected compared with
modified Urca. However, as we will show in the next section, at the temperatures 
characteristic of neutron star mergers this is not the case.

We note that a similar analysis of the Fermi-Dirac factors can be done with
the modified Urca process, but it simply reproduces the expected finding that
at any density the dominant contribution comes from particles close to their
Fermi surfaces, so the Fermi surface approximation is always valid for
modified Urca and there is never any Boltzmann suppression of the rate.  

In Sec.~\ref{sec:intro}, we estimated that 
at densities below the direct Urca threshold
the Boltzmann-suppressed direct Urca electron capture rate would match the modified Urca rate once the temperature rose to around $1$ or $2\,\MeV$.   As we will see in the next section, a full calculation
confirms this estimate, showing that at $T\gtrsim 1\,\MeV$ the
contribution of below-threshold direct Urca processes leads to corrections
to the low-temperature criterion for beta equilibrium.
Since the dominant contribution to the below-threshold direct Urca rates comes
from particles that are far from their Fermi surfaces, we now calculate the
direct Urca rates exactly, performing the entire momentum space integral.

%%%%%%%%%%%%%%%%%%%%%%%%%%%%%%%%%%%%
\section{Exact direct Urca calculation} 
\label{sec:exact}

Instead of assuming that all particles lie on their Fermi surfaces, we numerically evaluate the direct Urca rate integrals (\ref{eq:ndecay}) and (\ref{eq:ecapture}) with non-relativistic nucleons, but without any further approximation, allowing the particles to have any set of momenta that is consistent with energy-momentum conservation.  The details of the calculation, which reduces to a three dimensional numerical integral, are given 
in Appendix~\ref{sec:rate-integral}.

%%%%%%%%%%%%%%%%%%%%%%%%%%%%%%%%
\subsection{Urca rates}

Figs.~\ref{fig:dU_500keV} and \ref{fig:dU_5MeV} show 
in their upper panels
the rates of various Urca processes in APR nuclear matter that obeys the
low-temperature beta equilibrium criterion \eqn{eq:beta}.
As we will see, at $T\gtrsim 1\,\MeV$ the Fermi surface approximation
starts breaking down and Eq.~\eqn{eq:beta} is no longer the
correct criterion for beta equilibrium:
an additional isospin-coupled chemical potential is needed
to achieve beta equilibrium, and its magnitude $\mud$ is shown in the lower
panel.

%and approximate modified Urca rates in APR nuclear matter that obeys the
%standard beta equilibrium criterion \eqn{eq:beta} over a range of densities
%relevant for neutron stars, at temperatures of $500\,\text{keV}$and
%$10\,\MeV$.

In Fig.~\ref{fig:dU_500keV}, we see that at a temperature of $500 \,\keV$, the Fermi surface approximation is reasonably accurate.
Well above the direct Urca threshold density the neutron decay and electron capture rates are almost identical and agree well with the Fermi surface approximation, so
that when the low-temperature beta equilibrium criterion \eqn{eq:beta} is obeyed
the net rate of neutron or proton creation is zero. Consequently, 
to the accuracy of our calculation ($\mu_{\delta}$ is accurate to about $\pm 150\,\keV$, described in Sec.~\ref{sec:full}) there is
no need for any additional isospin chemical potential to enforce beta
equilibrium.
As the density drops below the threshold value, the direct Urca rates
drop below the modified Urca rate and become negligible, and the modified Urca rates for neutron decay and electron capture are identical so again there is no net creation of neutrons or protons, and no noticeable 
modification to the low-temperature beta equilibrium criterion.
However, it is interesting to note that
below (and even slightly above) threshold the direct Urca rates for neutron decay and electron capture are not the same. 
The deviation increases as the density goes further below threshold.  The size of the discrepancy agrees with our analysis in Fig \ref{fig:FS_deviation}, where we determine the exponential suppression of each direct Urca rate, due to the Fermi-Dirac factors.  Only right below threshold is there a region where the two direct Urca rates are different, but both are larger than the modified Urca rates, requiring a finite but small $\mud$ to establish true beta equilibrium.  As temperature decreases further, this effect will vanish, and the low-temperature beta equilibrium criterion (\ref{eq:beta}) will be increasingly valid.  

In Fig.~\ref{fig:dU_5MeV}
we see that when we increase the temperature to $T=5\,\MeV$
the Fermi surface approximation becomes unreliable.
The direct Urca electron capture rate agrees with the Fermi surface result
above threshold, but it also shows no suppression below threshold, dominating over
modified Urca and direct Urca neutron decay at all densities for which APR is well defined. This means that when $\mu_n=\mu_p+\mu_e$ (Eq. \ref{eq:beta}), there is a nonzero net rate of proton to neutron conversion, implying that
the system is not in beta equilibrium.

\begin{figure}
\includegraphics[width=\hsize]{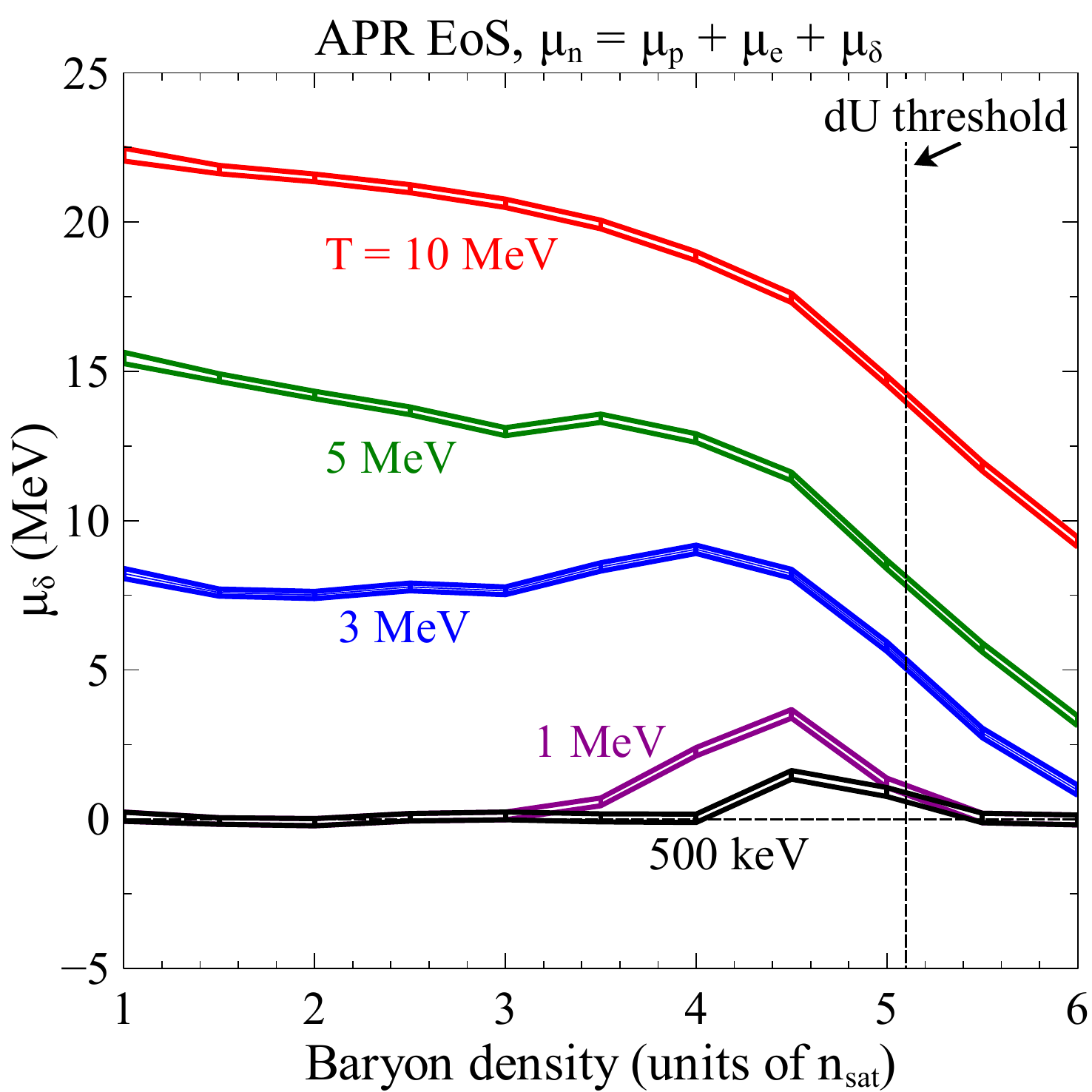}
\caption{
The isospin-coupled chemical potential $\mud$ needed to achieve true beta equilibrium in APR matter at various temperatures.  For a given temperature, the upper and lower curves indicate the range of values of $\mu_{\delta}$ that are consistent with our estimates of the theoretical uncertainty.  Further details on the uncertainty are given in the text.}
\label{fig:dmu_plot}
\end{figure} 

%\subsection{General criterion for beta equilibrium}
%%%%%%%%%%%%%%%%%%%%%%%%%%%%%%%%%%
\subsection{Full criterion for beta equilibrium}
\label{sec:full}
The fact that electron capture is much less suppressed than neutron decay at $T\gtrsim 1\,\MeV$ means that the system will be driven away from the state that obeys the standard low-temperature beta equilibrium criterion. The predominance of electron capture drives the neutron Fermi energy up and the proton Fermi energy down, effectively introducing an additional chemical potential that couples to the third component of isospin.

The general criterion for beta equilibrium is
\beq
\mu_n = \mu_p + \mu_e + \mud \ ,
\label{eq:beta-general}
\eeq
where $\mud = -\muI$ where $\muI$ is the chemical potential that
couples to isospin, normalized according to a convention \cite{isospin_convention} where the proton has isospin $+1/2$ and the neutron
has isospin $-1/2$.
% mu_n = mu_B - 1 mu_I
% mu_p = mu_B - mu_e + 1 mu_I
% So mu_n = mu_p + mu+e - 2 mu_I

As the proton density drops, the rates of electron capture and neutron decay move towards each other and eventually balance when $\mud$ reaches its equilibrium value, which depends on the density and the temperature. This value of $\mu_{\delta}$ at $T = 5\,\MeV$ is shown in the bottom panel of Fig.~\ref{fig:dU_5MeV}.  We see that below the direct Urca threshold $\mud$ is about 15\,\MeV\ and it decreases but remains non-negligible above threshold as well.

In Fig.~\ref{fig:dmu_plot}, we show, for several temperatures, the magnitude of the additional isospin-coupled chemical potential $\mud$ needed to achieve true beta equilibrium.  At low temperatures, below $1\,\MeV$, the standard criterion Eq.~\eqn{eq:beta} is correct to within about 1\,\MeV, the only noticeable correction, $\mu_{\delta} \approx 1.5 \,\MeV$, occurring right below threshold where direct Urca electron capture begins to dominate over modified Urca. After that, however, the correction term rises quickly with temperature: at $T=5\,\MeV$ we need $\mud\sim 15\,\MeV$ and at $T=10\,\MeV$ we need $\mud\sim 23\,\MeV$.  Although $\mud$ drops with density once we reach the direct Urca threshold, that decrease becomes quite slow at these higher temperatures.  If the neutrino trapping temperature is indeed around $5\, \MeV$ (see Appendix \ref{sec:MFP}), then our calculations of $\mu_{\delta}$ are only physically relevant at temperatures below $5\, \MeV$.  Above the neutrino trapping temperature we expect that neutrinos are in statistical equilibrium with a  chemical potential $\mu_\nu$ obeying the detailed balance relation $\mu_n + \mu_\nu = \mu_p + \mu_e$.

The APR equation of state is based on variational calculations of the energy (as a function of density) of pure neutron matter (PNM, $x_p = 0$) and symmetric nuclear matter (SNM, $x_p = 0.5$).  An interpolation scheme was used to get to the intermediate proton fraction, for a given density, that satisfies the low-temperature beta equilibrium condition (\ref{eq:beta}) \cite{APR}.  Thus, all thermodynamic quantities calculated within the APR framework are functions of both baryon density and proton fraction.  To find the value of $\mu_{\delta} = \mu_n-\mu_p-\mu_e$ necessary to achieve true beta equilibrium at a given density and temperature, we varied the proton fraction in discrete steps until we found a proton fraction that at which there was net neutron production, and an adjacent-step proton fraction at which there was net neutron destruction, giving us an upper and lower bound on $\mu_{\delta}$. These upper and lower bounds provide the theoretical error on $\mu_{\delta}$.  The uncertainty in the direct Urca rate calculations, discussed at the end of Appendix \ref{sec:rate-integral}, is smaller than the binning of $\mu_{\delta}$, which is $\approx 300\,\keV$.  

\begin{figure}
\includegraphics[width=\hsize]{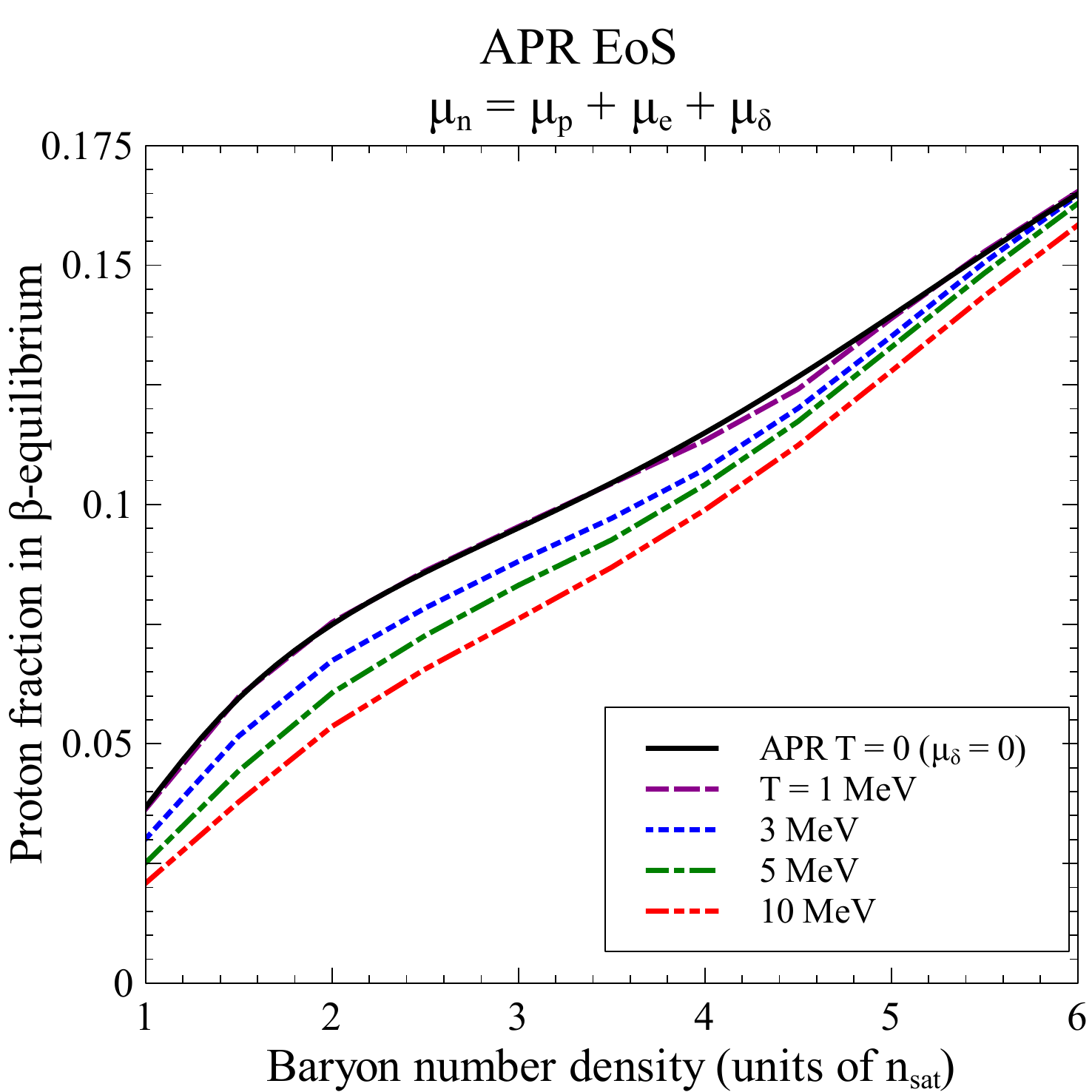}
\caption{Proton fraction in neutrino-transparent nuclear matter in true beta equilibrium, for several different temperatures.  As temperature increases, the beta-equilibrated nuclear matter becomes more neutron-rich than predicted by the low-temperature beta equilibrium condition.}
\label{fig:new_xp}
\end{figure}

The temperature-dependent correction $\mud$ to the beta equilibrium condition arises from the temperature dependence of the proton fraction in true beta equilibrium. In Fig.~\ref{fig:new_xp} we plot the proton fraction in true beta equilibrium as a function of density, for various temperatures.  As temperature rises above $1\,\MeV$, the proton fraction drops. This reflects the predominance of electron capture over neutron decay seen in Fig.~\ref{fig:dU_5MeV}.

\begin{figure}
\includegraphics[width=\hsize]{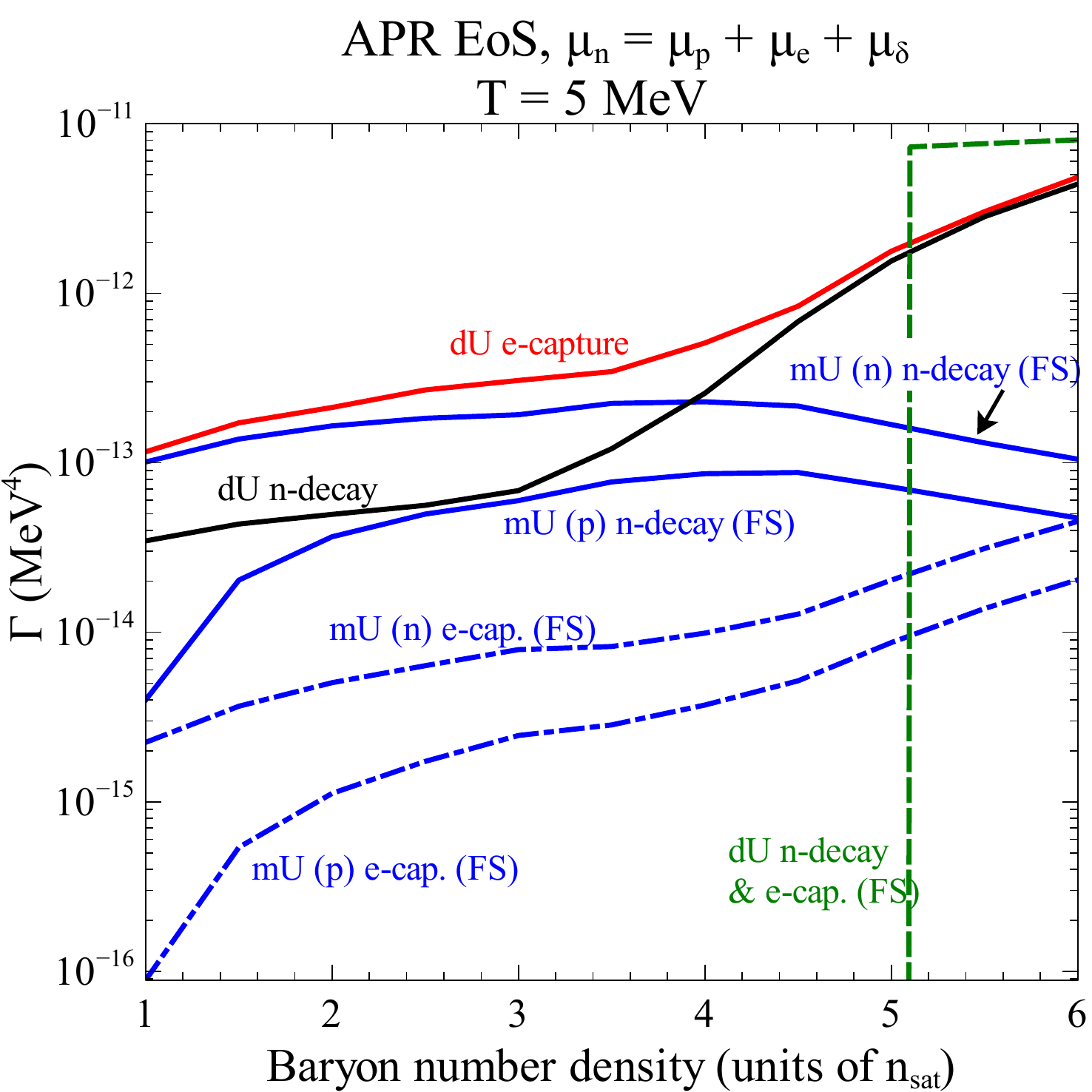}
\caption{
Urca rates in true beta equilibrium, $\mu_n = \mu_p + \mu_e + \mud$.  Above threshold, the two direct Urca processes dominate.  Below threshold, direct Urca electron capture balances against modified Urca neutron decay (n-spectator) and, to a lesser extent, direct Urca neutron decay.
}
\label{fig:beta-true}
\end{figure}

\begin{figure}
\includegraphics[width=\hsize]{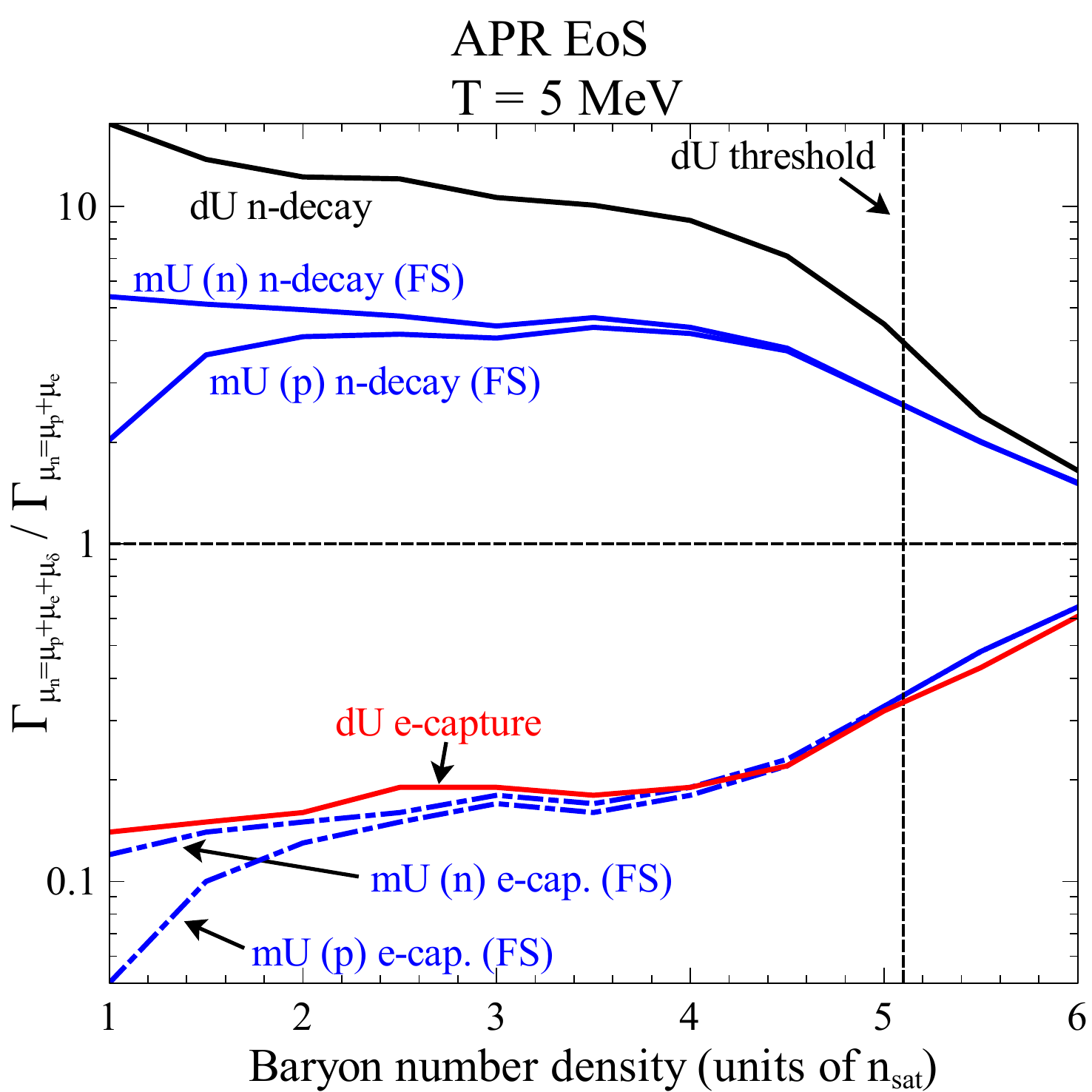}
\caption{Fractional change in the six Urca rates at $T = 5\,\MeV$ when we change $\mud$ from zero to the value for true beta equilibrium.  Below threshold, the change is the most prominent, and as density increases to far above the threshold density, the true beta equilibrium condition approaches the behavior of the low-temperature beta equilibrium condition (\ref{eq:beta}).  
}
\label{fig:frac_change}
\end{figure}

%%%%%%%%%%%%%%%%%%%%%%%%%%%%%%%%
\section{Conclusions}

We have shown that that the standard low-temperature criterion
\eqn{eq:beta} for beta equilibrium 
breaks down in neutrino-transparent nuclear matter
at densities above nuclear saturation density 
and temperatures above about 1\,\MeV. An additional isospin
chemical potential \eqn{eq:beta-general} is required to obtain
true equilibrium under the weak interactions.
The ultimate reason for this is that neutrinos are not
in thermal equilibrium, so the two reactions \eqn{eq:Urca} that
drive the system to equilibrium are not exact inverses of each other
(neutrinos can only occur in final states) so the principle of detailed
balance does not apply. 
Our calculations for nuclear matter
obeying the APR equation of state show (Fig.~\ref{fig:dmu_plot})
that the isospin chemical potential $\mud = \mu_n-\mu_p-\mu_e$
required to obtain beta equilibrium
becomes greater than about 5\,\MeV\ as the temperature rises above
1\,\MeV, and reaches a maximum value around 23\,\MeV\ at temperatures 
of 10\,\MeV.

We have recalculated the Urca rates for APR nuclear matter at
$T=5\,\MeV$ in true beta equilibrium.  The results are given in Fig.~\ref{fig:beta-true}, which shows that there are three processes that play the central role in beta equilibration. Above threshold, direct Urca neutron decay balances with direct Urca electron capture.  Far below threshold, neutron-spectator modified Urca neutron decay competes with direct Urca electron capture, but as threshold is approached from below, direct Urca neutron decay becomes increasingly important, eventually becoming more important than neutron-spectator modified Urca neutron decay.

In Fig.~\ref{fig:frac_change}, we show the fractional change in the six direct Urca rates when the correct beta equilibrium condition is imposed, compared with the low-temperature criterion for beta equilibrium (\ref{eq:beta}).  Far below threshold, the Urca rates increase or decrease by a factor of $10$, and far above threshold, the Urca rates approach their low-temperature beta equilibrium (\ref{eq:beta}) values. 

The adjustment $\mud$ to the low-temperature criterion for beta equilibrium has implications for any effect stemming from weak interactions in beta equilibrated nuclear matter, in the temperature range from about $1\,\MeV$ to the neutrino trapping temperature.  For example, the nuclear matter in neutron star mergers, which certainly reaches such temperatures, undergoes significant density oscillations during the merger.  The density oscillations may be damped by bulk viscosity, which arises from a resonance between the density oscillations and beta equilibration of the proton fraction, which occurs via Urca processes  \cite{Alford_transport,alford_bulk_viscosity,Yakovlev_review}.  The calculation of bulk viscosity relies on examining the adjustment of the neutron decay and electron capture rates as the nuclear matter is pushed out of beta equilibrium, so if the criterion for beta equilibrium is modified, it is likely that this will affect the calculated bulk viscosity for material in neutron star mergers.

While we used the APR equation of state, which has two and three body nuclear interactions, to describe the nuclear matter in which the Urca processes take place, our calculation used the conventional Fermi-Dirac factors which assume that single-nucleon momentum eigenstates are also energy eigenstates.  However, it is known \cite{Dickhoff_1} that short-range nucleon-nucleon interactions lead to a depletion of low-momentum states in the nucleon Fermi sea and a high momentum tail above the Fermi surface, so a truly consistent calculation of the Urca rates would use modified Fermi-Dirac factors.  Additionally, the fact that an energy eigenstate actually corresponds to a superposition of multi-particle/hole eigenstates\cite{particle_hole_excitations} could further blur the direct Urca threshold beyond the finite temperature effects that we discuss in this paper.  

In this work we did not consider particle processes involving muons, although the APR equation of state includes contributions from muons.  Muons can participate in Urca processes and also in leptonic processes such as $\mu^- \rightarrow e^-\ +\ \bar{\nu}_e\ +\ \nu_{\mu}$.  A complete treatment of the Urca processes in neutrino-transparent nuclear matter would introduce another chemical potential $\mu_{f}$ (which couples to lepton flavor, differentiating electrons from muons) whose value at a given temperature and density is determined by balancing the six muon Urca process rates.

\begin{acknowledgments}
We thank Andreas Windisch, Kai Schwenzer, Wim Dickhoff, Micaela Oertel, Peter Shternin, and Luke Roberts for discussions.  This research was partly supported by the U.S. Department of Energy, 
 Office of Science, Office of Nuclear Physics, under Award No.~\#DE-FG02-05ER41375. 
\end{acknowledgments}
%%%%%%%%%%%%%%%%%%%%%%%%%%%%%%%%%%%%%%%%%%%%%%%%%%%

\appendix
%%%%%%%%%%%%%%%%%%%%%%%%%%%%
\section{Neutrino mean free path}
\label{sec:MFP}

Our calculations are applicable to matter in which neutrinos are not trapped, which means our results
are valid at temperatures up to the neutrino trapping temperature for a neutron star. 
This is determined by the length of the
neutrino mean free path compared with the size of the star. The mean free path depends on the neutrino energy and the temperature, density, and composition of the nuclear matter with which it interacts.  To get a reasonable idea of the neutrino trapping temperature we used a start-of-the-art code developed in Refs.~\cite{roberts_reddy,nuclear_mean_field}, and available online in \cite{roberts_reddy_code}, to calculate the neutrino mean free paths due to charged and neutral current reactions in nuclear matter above nuclear saturation density.  We used proton fractions and nuclear mean field values from the HS(DD2) equation of state\cite{Hempel:2009mc,Typel:2009sy}, which is tabulated online in \cite{Hempel_website}. The HS(DD2) equation of state is a relativistic mean field theory consistent with most data from nuclear experiments and astrophysical constraints \cite{Fischer:2013eka,Oertel:2016bki}. Unlike APR, it is tabulated for finite temperatures, making it suitable for estimating the temperature dependence of the mean free path.

\begin{figure}
\includegraphics[width=0.85\hsize]{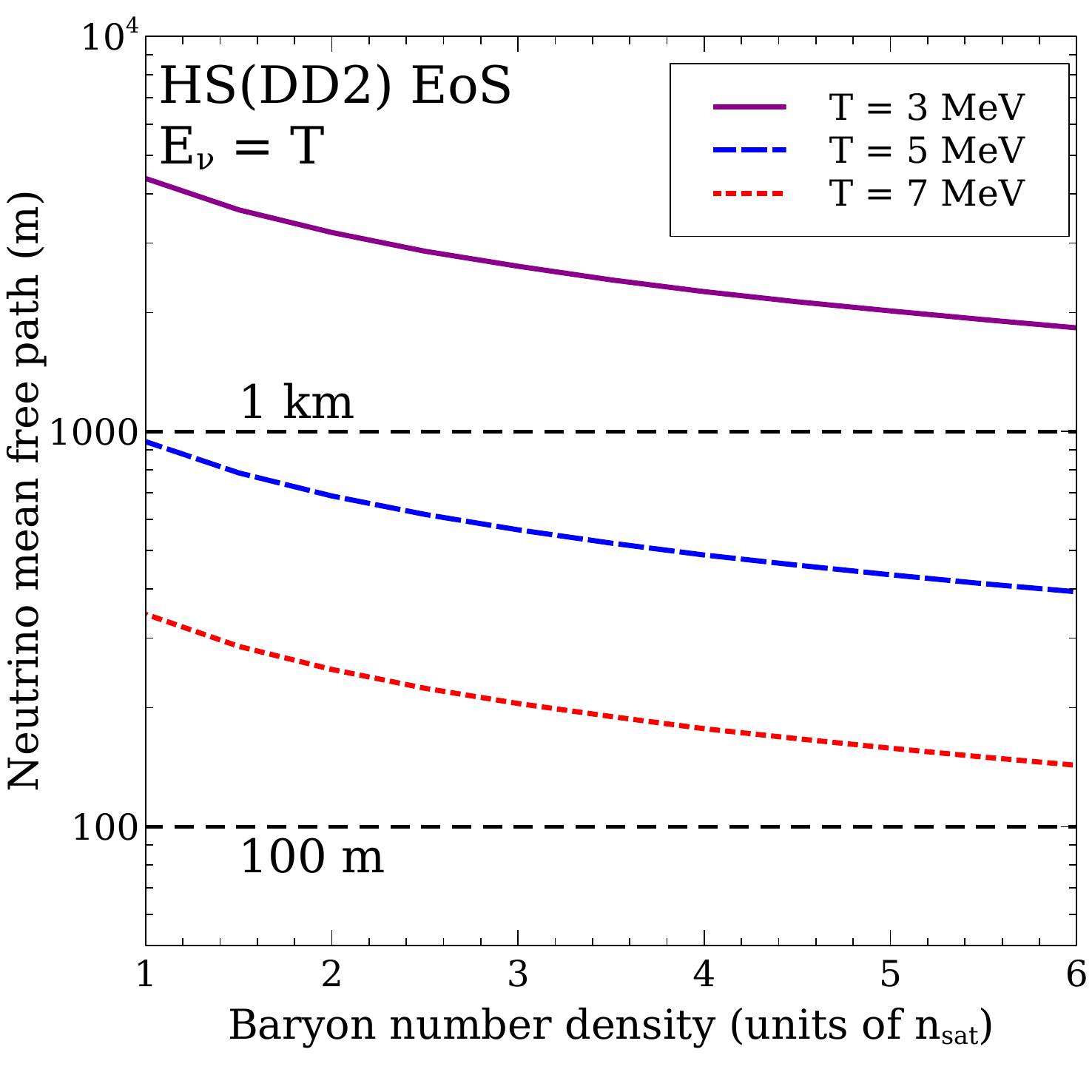}
\caption{Mean free path of neutrinos in nuclear matter described by the HS(DD2) equation of state (EoS). As temperature rises above 5\,MeV it becomes more likely that neutrinos will be trapped inside the merger region.
}
\label{fig:mfp}
\end{figure}

 The mean free path for neutrinos with energy equal to the temperature is plotted in Fig.~\ref{fig:mfp}.  We show the mean free path for neutral-current neutrino-neutron scattering, since that is the dominant scattering process at densities above  nuclear saturation density.  At a temperature of 3 MeV, across the relevant range of densities, neutrinos always have a mean free path of several kilometers, so they easily escape the merger region (whose radius is $\sim 10\,\mbox{km}$).  At $T=5\,\MeV$, the mean free path varies from about 1\,km at nuclear density density down to 0.5\,km at several times nuclear density,  and at $T=7\,\MeV$, the mean free path is typically a few hundred meters.
We conclude that it is reasonable to expect that neutrino trapping will only begin to become important when temperatures rise above 5 to 10\,MeV.

%%%%%%%%%%%%%%%%%%%%%%%%%%%%%%%%
\section{Modified Urca rates when $\mu_n \neq \mu_p + \mu_e$}
\label{sec:out_of_equil_mU}
We present here the Fermi-surface approximation of the modified Urca rates, when allowed to deviate from the low-temperature beta equilibrium criterion (\ref{eq:beta}) by an amount $\xi \equiv \left(\mu_n-\mu_p-\mu_e\right)/T$.  In the following expressions for the rates the non-equilibrium behavior is encapsulated in the function
\begin{align}
&F(\xi) = -\left(\xi^4+10\pi^2\xi^2+9\pi^4\right)\text{Li}_3(-e^{\xi}) \\
&+12\xi\left(\xi^2+5\pi^2\right)\text{Li}_4(-e^{\xi}) -24\left(3\xi^2+5\pi^2\right)\text{Li}_5(-e^{\xi})\nonumber\\
&+240\xi \text{Li}_6(-e^{\xi})-360 \text{Li}_7(-e^{\xi})\nonumber,
\end{align}
where $\text{Li}_n$ is the Polylogarithm function of order $n$.  We note that $F(0) \approx 2300.$
The rate of modified Urca neutron decay with a neutron spectator is
\begin{align}
\Gamma_{mU, nd (n)}(\xi) &= \dfrac{7}{64\pi^9} G^2 g_A^2 f_{\pi NN}^4 \dfrac{m_n^3m_p}{m_{\pi}^4}\\
&\times\dfrac{p_{F_n}^4p_{F_p}}{\left(p_{F_n}^2+m_{\pi}^2\right)^2} F(\xi) \vartheta_n T^7\nonumber
\end{align}
where $\vartheta_n$ is defined as in Eq.~\eqn{eq:mUrca_n}.  The rate of modified Urca electron capture with a neutron spectator is 
\begin{equation}
\Gamma_{mU, ec (n)}(\xi) = \Gamma_{mU, nd (n)}(-\xi),
\end{equation}
 and so the two neutron-spectator modified Urca rates agree in low-temperature beta equilibrium (\ref{eq:beta}).
The Fermi surface approximation of the modified Urca neutron decay process with a proton spectator is
\begin{align}
&\Gamma_{mU, nd (p)}(\xi) = \dfrac{1}{64\pi^9}G^2g_A^2f_{\pi NN}^4 \dfrac{m_nm_p^3}{m_{\pi}^4}\\
&\times \dfrac{p_{F_n}\left(p_{F_n}-p_{F_p}\right)^4}{\left(\left(p_{F_n}-p_{F_p}\right)^2+m_{\pi}^2\right)^2}F(\xi)\vartheta_p T^7\nonumber
\end{align}
with $\vartheta_p$ defined as in Eq.~\eqn{eq:mUrca_p}, and the modified Urca electron capture rate with a proton spectator is
\begin{equation}
\Gamma_{mU, ec (p)}(\xi) = \Gamma_{mU, nd (p)}(-\xi),
\end{equation}
where again both modified Urca rates with a proton spectator agree in low-temperature beta equilibrium (\ref{eq:beta}).

%%%%%%%%%%%%%%%%%%%%%%%%%%%%%%%%

\vspace{4ex}
\section{Exact direct Urca rate integral}
\label{sec:rate-integral}

The neutron decay rate given by a twelve dimensional integral in Eq.~\eqn{eq:ndecay} can be reduced, without approximation, to a three dimensional integral.  Integrating over neutrino three-momentum, we have
\begin{align}
\Gamma_{n} &=\dfrac{G^2}{128\pi^8} \int \mathop{d^3p_n}\mathop{d^3p_p}\mathop{d^3p_e}f_n\left(1-f_p\right)\left(1-f_e\right)\nonumber\\
&\times\delta(q-\vert \mathbf{p}_n-\mathbf{p}_p-\mathbf{p}_e\vert )\\
&\times \left( 1+3g_A^2+\left(1-g_A^2\right)\hat{\mathbf{p}}_e \cdot \dfrac{\mathbf{p}_n-\mathbf{p}_p-\mathbf{p}_e}{\vert \mathbf{p}_n-\mathbf{p}_p-\mathbf{p}_e\vert}\right),\nonumber
\end{align}
where we define $q \equiv E_n - E_p - E_e$, and the ``hat'' denotes a unit vector.
We adopt spherical coordinates for the momentum of each of the three particles.  We have the freedom to choose the coordinates such that the neutron momentum lies along the $z$ axis, and the proton momentum lies in the same plane as the neutron momentum, and so the momentum unit vectors are written as
\begin{align}
\hat{\mathbf{p}}_n &= \left(0,0,1\right)\\
\hat{\mathbf{p}}_p &= \left(\sqrt{1-z_p^2},0,z_p\right)\\
\hat{\mathbf{p}}_e &= \left(\sqrt{1-z_e^2}\cos{\phi},\sqrt{1-z_e^2}\sin{\phi},z_e\right),
\end{align}
where $z_p$ and $z_e$ are cosines of the polar angles of the proton and electron momenta and as such, take values from $-1$ to $1$.  The azimuthal angle $\phi$ of the electron with respect to the plane formed by the proton and neutron momenta ranges from $0$ to $2\pi$.

This choice of coordinates allows us to integrate over the three trivial angles, giving a factor of $8\pi^2$.  The rate integral can now be written as
\begin{align}
\Gamma_{n} &= \dfrac{G^2}{16\pi^6}\int_0^{\infty} \mathop{dp_n}\mathop{dp_p}\mathop{dp_e} p_n^2p_p^2p_e^2 \\
&\times f_n\left(1-f_p\right)\left(1-f_e\right)  I(p_n,p_p,p_e),\nonumber
\end{align}
where 
\begin{widetext}
\begin{align}
I &= \int_{-1}^1\mathop{dz_e}\int_{-1}^1\mathop{dz_p}\int_0^{2\pi}\mathop{d\phi}\delta\Bigg(q-\sqrt{p_n^2+p_p^2+p_e^2+2p_pp_e\sqrt{1-z_p^2}\sqrt{1-z_e^2}\cos{\phi}+2p_pp_ez_pz_e-2p_np_pz_p-2p_np_ez_e   }\Bigg)\\
&\times \left( 1+3g_A^2+\left(1-g_A^2\right)\dfrac{p_nz_e-p_pz_pz_e-p_e-p_p\sqrt{1-z_p^2}\sqrt{1-z_e^2}\cos{\phi}}{\sqrt{ p_n^2+p_p^2+p_e^2+2p_pp_e\sqrt{1-z_p^2}\sqrt{1-z_e^2}\cos{\phi}+2p_pp_ez_pz_e-2p_np_pz_p-2p_np_ez_e  }} \right).\nonumber
\end{align}
\end{widetext}
We do the $\phi$ integral first, using the delta function.  Clearly we require $q>0$, because if $q$ is negative, the argument of the delta function could never be zero and thus the integral would be zero.  For $q>0$, the delta function argument vanishes for either zero or two values of $\phi$ between $0$ and $2\pi$.  We find that
\begin{widetext}
\begin{align}
I &= 4\vert q\vert\int_{-1}^1\mathop{dz_p}\left(1+3g_A^2+\dfrac{1-g_A^2}{2p_eq}\left(p_n^2+p_p^2-p_e^2-q^2-2p_np_pz_p\right)\right)\Theta(q)\\
&\times \int_{-1}^{1}\mathop{dz_e} \dfrac{1}{\sqrt{ 4p_p^2p_e^2\left(1-z_p^2\right)\left(1-z_e^2\right)-\left(q^2-p_n^2-p_p^2-p_e^2-2p_pp_ez_pz_e+2p_np_pz_p+2p_np_ez_e\right)^2 }}\Theta(B),\nonumber
\end{align}
\end{widetext}
where the step function $\Theta(B)$ enforces $B>0$, where 
\begin{align}
&B = 2p_pp_e\sqrt{1-z_p^2}\sqrt{1-z_e^2}\\
&-\vert q^2-p_n^2-p_p^2-p_e^2-2p_pp_ez_pz_e+2p_np_pz_p+2p_np_ez_e\vert.\nonumber
\end{align}
This is the condition for there to be two, not zero, values of $\phi$ in the integration range which make the argument of the delta function vanish and thus contribute to the integral.

We now evaluate the $z_e$ integral, noting that the step function $\Theta(B)$ adjusts the range of integration.  
Only if $C>0$, where
\begin{equation}
C = 2p_e\vert q\vert-\vert p_e^2+q^2-p_n^2-p_p^2+2p_np_pz_p\vert,
\end{equation}
is the step function nonzero for any range of $z_e$ in the interval $[-1,1]$, in which case the step function is nonzero only for $z_e^- < z_e < z_e^+$, where $z_e^{\pm}$ lie inside the interval $[-1,1]$, and thus $z_e^{\pm}$ become the new integration bounds.  Doing the $z_e$ integral, we find
\begin{align}
&I = \dfrac{2\pi \vert q\vert}{p_e}\Theta(q)\int_{-1}^1\mathop{dz_p}\Theta(C)\\
&\times \left( \dfrac{1+3g_A^2+\dfrac{1-g_A^2}{2p_eq}\left(p_n^2+p_p^2-p_e^2-q^2-2p_np_pz_p\right)}{\sqrt{p_n^2+p_p^2-2p_np_pz_p}}\right)\nonumber.
\end{align}

The step function $\Theta(C)$ creates a restriction on the bounds of $z_p$ and so the actual range of integration over $z_p$ is the intersection of the intervals $[-1,1]$ and $[z_p^-,z_p^+]$, where $z_p^{\pm} = (p_n^2+p_p^2-p_e^2-q^2 \pm 2p_e\vert q \vert )/(2p_n p_p)$, and so the range of integration will depend on the values of  $\{p_n,p_p,p_e\}$.  Evaluating the integral over $z_p$ with the bounds $y^+$ and $y^-$, we have
\begin{equation}
I(p_n,p_p,p_e) =  \frac{2\pi\vert q\vert}{p_np_pp_e}\Theta(q) J(p_n,p_p,p_e)
\end{equation}
where
\begin{align}
\label{eq:Jdef}
J(p_n,p_p,p_e) &= \bigg[ \Big(1+3g_A^2- (1-g_A^2)\frac{p_e^2+q^2}{2p_eq}\Big)y^{1/2} \\
&+\frac{1-g_A^2}{6p_eq}y^{3/2}  \bigg] \Bigg\rvert_{y=y^-}^{y=y^+} \nonumber
\end{align}
with
\begin{equation}
\label{eq:upper}
y^+ =
 \begin{cases}
(p_e+\vert q\vert)^2 & \text{ if } -1<z_p^-<1<z_p^+ \\
(p_e+\vert q\vert)^2 & \text{ if } -1<z_p^-<z_p^+<1 \\
(p_n+p_p)^2 & \text{ if  }  z_p^-<-1<1<z_p^+\\
(p_n+p_p)^2 &  \text{ if  } z_p^-<-1<z_p^+<1
\end{cases}
\end{equation}
and
\begin{equation}
\label{eq:lower}
y^- = 
\begin{cases}
(p_n-p_p)^2 & \text{ if } -1<z_p^-<1<z_p^+\\
(p_e-\vert q \vert)^2 & \text{ if } -1 < z_p^- < z_p^+<1\\
(p_n-p_p)^2 & \text{ if } z_p^- < -1 < 1 < z_p^+\\
(p_e-\vert q\vert)^2 & \text{ if } z_p^- < -1 < z_p^+ < 1.
\end{cases}
\end{equation}

Thus, the direct Urca neutron decay rate is
\begin{align}
\label{eq:final_urca_rate}
\Gamma_n &= \dfrac{G^2}{8\pi^5}\int_0^{\infty} \mathop{dp_n}\mathop{dp_p}\mathop{dp_e} p_np_pp_e\vert q\vert \Theta(q)\\
&\times f_n\left(1-f_p\right)\left(1-f_e\right)  J(p_n,p_p,p_e) \nonumber
\end{align}
with $J(p_n,p_p,p_e)$ as defined in Eqs.~\ref{eq:Jdef}, \ref{eq:upper}, and \ref{eq:lower}.

The direct Urca electron capture rate integral is identical, except for $f_n\left(1-f_p\right)\left(1-f_e\right)$ is replaced by $\left(1-f_n\right)f_pf_e$ and, because the neutrino is now on the same side of the reaction as the neutron, instead of with the electron and proton, $\Theta(q)$ is replaced by $\Theta(-q)$.

 The remaining three dimensional integral in Eq.~\ref{eq:final_urca_rate} can be done numerically (we used Mathematica's Monte Carlo integration routine), giving the direct Urca rate results shown in Figs.~\ref{fig:dU_500keV} and \ref{fig:dU_5MeV}.  Mathematica's estimated error on the numerical integrals is under twenty percent, and repeated evaluation of the integrals leads to results within ten percent of their average. 

%%%%%%%%%%%%%%%%%%%%%%%%%%%%%%%%%%%%%%

\bibliographystyle{JHEP}
\bibliography{beta_equilibrium}{}

\end{document}